\documentclass{article}

    \usepackage[preprint]{neurips_2025}

\usepackage[utf8]{inputenc} % allow utf-8 input
\usepackage[T1]{fontenc}    % use 8-bit T1 fonts
\usepackage{hyperref}       % hyperlinks
\usepackage{url}            % simple URL typesetting
\usepackage{booktabs}       % professional-quality tables
\usepackage{amsfonts}       % blackboard math symbols
\usepackage{natbib}         % for \citep and citations in captions
\usepackage{nicefrac}       % compact symbols for 1/2, etc.
\usepackage{microtype}      % microtypography
\usepackage{xcolor}         % colors
\usepackage{graphicx}

\usepackage{times}
\usepackage{latexsym}
\usepackage{inconsolata}

\usepackage{amsmath}
\usepackage{multirow}
\usepackage{caption}
\usepackage{subcaption}
\usepackage{algorithm}
\usepackage{algorithmic}
\usepackage{longtable}
\usepackage{amssymb}
\usepackage{pifont}
\usepackage{tabularx}
\usepackage{bbding}
\usepackage{utfsym}
\usepackage{hyperref}
\usepackage{enumitem}
\usepackage{setspace}
\usepackage{courier} % light font weight
\usepackage{url}            % simple URL typesetting
\usepackage{xspace}
\usepackage{wrapfig}
\usepackage{graphicx}
\usepackage{lipsum}

\usepackage{array}        % for \newcolumntype
\usepackage{tabularx}      % for X‐columns
\usepackage{booktabs}      % for \toprule, \midrule, \bottomrule
\usepackage{colortbl} % for \definecolor and \rowcolors
\usepackage{caption}       % to tune caption widths
\usepackage{arydshln}
\usepackage{amssymb} % for \checkmark
\usepackage{pifont} % for \ding{55}
\definecolor{mydarkblue}{rgb}{0,0.4,0.8} 
\definecolor{mylightblue}{rgb}{0.5,0.75,1}
\definecolor{NavyBlue}{HTML}{000080}
\hypersetup{
    colorlinks=true,
    linkcolor=red,
    citecolor=mydarkblue,
    filecolor=magenta,      
    urlcolor=magenta,
}

\usepackage{threeparttable}

\title{From Web Search towards Agentic Deep \underline{Research}: Incentivizing \underline{Search} with \underline{Re}asoning Agents}

\author{
  Weizhi Zhang\textsuperscript{1\thanks{Authors contributed equally to this research.} \hspace{0.01em} \thanks{Corresponding author: \texttt{wzhan42@uic.edu}}}, Yangning Li\textsuperscript{2\footnotemark[1]},
  Yuanchen Bei\textsuperscript{3\footnotemark[1]}, Junyu Luo\textsuperscript{4}, Guancheng Wan\textsuperscript{5}, Liangwei Yang\textsuperscript{6},\\
  \textbf{Chenxuan Xie\textsuperscript{7}, Yuyao Yang\textsuperscript{1}, Wei-Chieh Huang\textsuperscript{1}, Chunyu Miao\textsuperscript{1}, Henry Peng Zou\textsuperscript{1}, Xiao Luo\textsuperscript{5}, }\\
  \textbf{Yusheng Zhao\textsuperscript{4}, Yankai Chen\textsuperscript{1}, Chunkit Chan\textsuperscript{8}, Peilin Zhou\textsuperscript{9}, Xinyang Zhang\textsuperscript{10\thanks{Work does note related the author's position at Amazon}}, Chenwei Zhang\textsuperscript{10\footnotemark[3]},} \\
  \textbf{ Jingbo Shang\textsuperscript{11}, Ming Zhang\textsuperscript{4}, Yangqiu Song\textsuperscript{8}, Irwin King\textsuperscript{12}, Philip S. Yu\textsuperscript{1}} \\
  \textbf{}\\
  \textsuperscript{1} University of Illinois Chicago, 
  \textsuperscript{2} Tsinghua University, 
  \textsuperscript{3} University of Illinois Urbana-Champaign, \\
  \textsuperscript{4} Peking University,
  \textsuperscript{5} University of California, Los Angeles,
  \textsuperscript{6} Salesforce AI Research, \\
  \textsuperscript{7} Zhejiang University of Technology,
  \textsuperscript{8} The Hong Kong University of Science and Technology, \\
  \textsuperscript{9} The Hong Kong University of Science and Technology (Guangzhou),
  \textsuperscript{10} Amazon,\\
  \textsuperscript{11} University of California, San Diego,
  \textsuperscript{12} The Chinese University of Hong Kong\\
}

\begin{document}

\maketitle

\begin{abstract}
Information retrieval is a cornerstone of modern knowledge acquisition, enabling billions of queries each day across diverse domains. However, traditional keyword-based search engines are increasingly inadequate for handling complex, multi-step information needs. Our position is that Large Language Models (LLMs), endowed with reasoning and agentic capabilities, are ushering in a new paradigm termed Agentic Deep Research. These systems transcend conventional information search techniques by tightly integrating autonomous reasoning, iterative retrieval, and information synthesis into a dynamic feedback loop. We trace the evolution from static web search to interactive, agent-based systems that plan, explore, and learn. We also introduce a test-time scaling law to formalize the impact of computational depth on reasoning and search. Supported by benchmark results and the rise of open-source implementations, we demonstrate that Agentic Deep Research not only significantly outperforms existing approaches, but is also poised to become the dominant paradigm for future information seeking. 
All the related resources, including industry products, research papers, benchmark datasets, and open-source implementations, are collected for the community in \url{https://github.com/DavidZWZ/Awesome-Deep-Research}.
\end{abstract}

\section{Introduction}

\begin{flushleft}

\textit{
“Introducing deep research: An agent that uses reasoning to synthesize large amounts of online information and complete multi-step research tasks for you.”}

\hfill -- \textit{OpenAI}
\end{flushleft}

Every day, billions of people search for information online ~\citep{amendola2023social}, and rely heavily on these online resources to make decisions across personal, professional, and societal contexts~\citep{zhang2024usimagent}. For decades, traditional web search engines based on keyword matching have served as the primary gateway to digital information. While once revolutionary, these systems increasingly struggle with complex, multi-faceted queries that demand nuanced understanding and synthesis~\citep{mo2024surveyconversationalsearch}. The growing inadequacy highlights their inherent limitations in contextual comprehension and knowledge integration.

Against this backdrop, information seeking and synthesis are undergoing a profound transformation with large language models (LLMs). Rather than merely enhancing traditional search paradigms, LLMs are poised to fundamentally replace them in addressing complex information needs. 
Initially, we witnessed the rise of LLMs as standalone, knowledgeable chatbots, which challenged the dominance of web search by offering more direct answers and a degree of synthesis, thereby reducing the user's burden of sifting through numerous links~\citep{Liu_2023}. However, single LLMs are tethered to static, offline knowledge. The subsequent integration of search and retrieval-augmented generation (RAG) marked a step forward, grounding LLMs in external data and mitigating issues like hallucination~\citep{ma2023query, yang2025cold}. Nevertheless, these naive RAG methods still struggle with real-world questions that require sophisticated multi-hop reasoning and strategic search planning, as they often cannot plan correct search paths for complex problems~\citep{yao2023tree}.

\begin{figure*}[t]
    \centering
    \includegraphics[width=\linewidth]{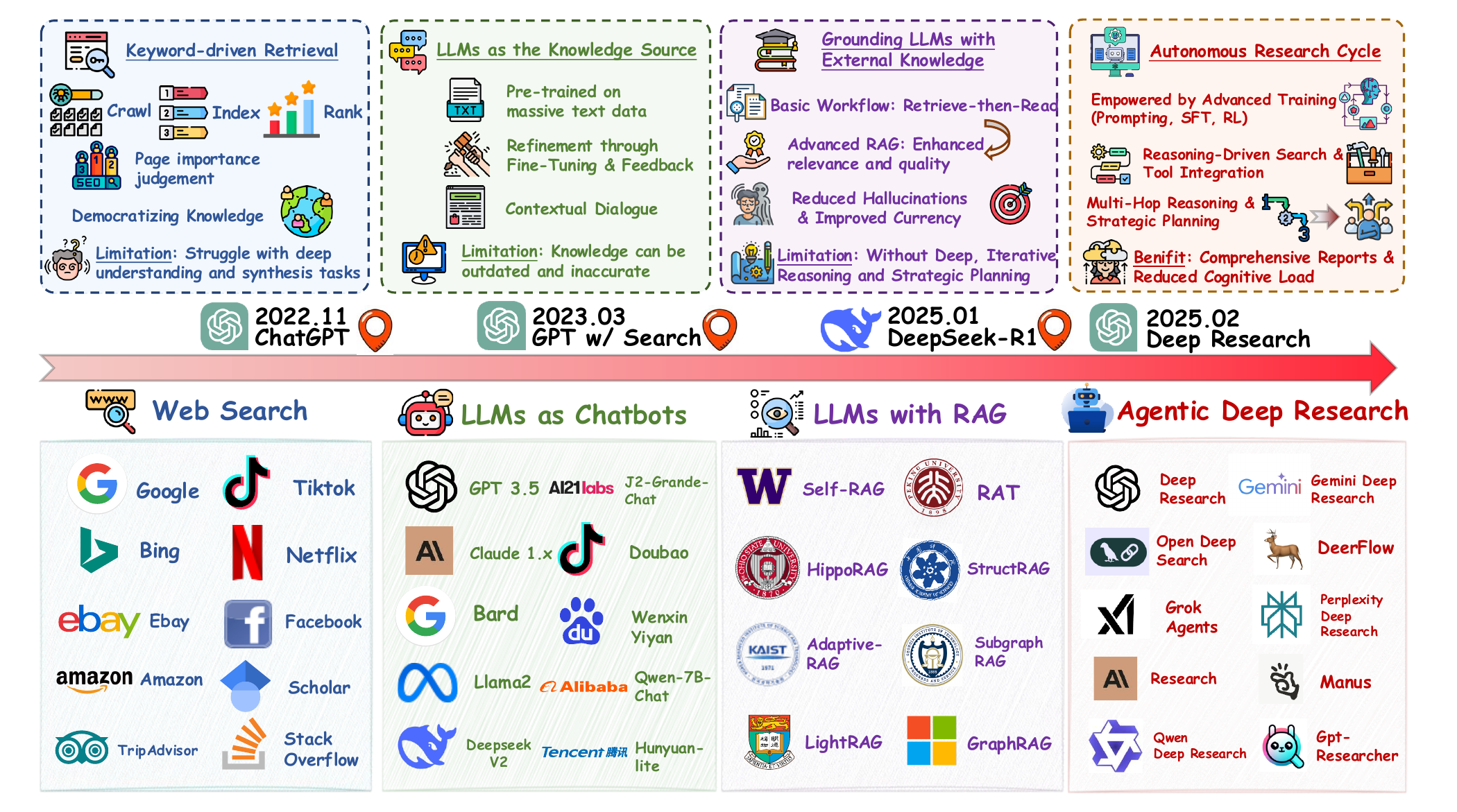}
    % \vspace{-0.5cm}
    \caption{The evolution of information search paradigms.}
    \label{fig: paradigm}
    \vspace{-0.5cm}
\end{figure*}

Recently, test-time scaling (TTS) has emerged as a potent paradigm for boosting the reasoning and agentic capabilities of LLMs~\citep{snell2024scaling}. It assigns additional computation during inference, enabling deeper problem-solving~\citep{zou2025testnuc, gu2025scaling}. 
Equipped with TTS on reasoning and search, LLMs are set to drive a new search paradigm termed \textbf{Agentic Deep \underline{Research}} systems, which are capable of autonomous \textbf{\underline{re}asoning}, on-demand \textbf{\underline{search}ing}, and iterative information synthesis.
Demonstrations from deep research products launched by OpenAI and Google highlight several key advantages of this paradigm: \textit{{(1) Comprehensive Understanding}}: Ability to dissect and address complex, multifaceted queries that overwhelm traditional methods~\citep{wei2022chain}; \textit{{(2) Enhanced Synthesis}}: Excels at synthesizing information from diverse, potentially conflicting sources into coherent and insightful narratives~\citep{cheng2025survey}; \textit{{(3) Reduced User Burden}}: Significantly decreases the cognitive load and manual effort required from users by automating laborious search steps~\citep{sami2024systematicliteraturereviewusing}.

Our position is that the LLM-driven Agentic Deep Research framework will inevitably become the dominant paradigm for future information search.
In this paper, we comprehensively investigate this paradigm shift and make four key contributions: (1) We systematically trace and analyze the evolutionary trajectory of information search paradigms, from traditional keyword-based search, through conversational LLM chatbots and naive search-augmented LLMs, and ultimately to the Agentic Deep Research; (2) We introduce the {test-time scaling (TTS) law for Deep Research}, a novel hypothesis formalizing the relationship between inference-time computational resources allocation and the resulting improvements in LLMs reasoning capabilities and knowledge depth; 
(3) We conduct extensive evaluations on existing Agentic Deep Research models, coupled with analyses of open-source implementations to support our position; and (4) As the first to holistically summarize the field of Deep Research, we offer a detailed exploration of critical future research directions, outlining both opportunities and challenges. Our discussion outlines a clear roadmap for this rapidly evolving field, underscoring how this evolution is fundamentally reshaping human interaction with real-world information and guiding further advancement on Agentic Deep Research.

% \clearpage

\section{Traditional Information Search Paradigms}
The evolution of search paradigms as in Figure~\ref{fig: paradigm} represents a fundamental transformation in how humans access and interact with information. This section examines three distinct frameworks that have shaped the landscape of information retrieval: traditional web search engines, Large Language Models (LLMs) as chatbots, and LLMs with Retrieval Augmented Generation (RAG) systems. Each paradigm offers unique capabilities and addresses prior information seeking challenges, from systematic web crawling and ranking to interactive dialogue and knowledge-augmented generation.

\subsection{Web Search}
Web search has fundamentally transformed information access in modern society, enabling near-instantaneous retrieval of knowledge that previously required days or months to locate~\citep{brin1998anatomy}. This revolutionary technology has democratized knowledge acquisition, accelerated economic development through improved information flow, and catalyzed scientific discovery by providing researchers with rapid access to cutting-edge developments. As the primary infrastructure for information retrieval in the Internet era, online  search engines like Google have continuously shaped how humans interact with the expanding information landscape~\citep{page1999pagerank}.

The applications of web search span diverse contexts, from general-purpose engines handling broad information needs to specialized platforms optimized for specific domains. General search engines utilize sophisticated algorithms to address multifaceted user queries~\citep{broder2002taxonomy}, while specialized systems like Google Scholar focus on academic literature, providing researchers with precise pathways to scholarly resources~\citep{noruzi2005google}. Content platforms (TikTok), social networks (Facebook), and e-commerce sites have developed internal search capabilities tailored to their unique content types and user behaviors, demonstrating how web search has permeated virtually every digital sphere. 

Basically, web search operates through three fundamental processes: crawling~\citep{khder2021web}, indexing~\citep{hendriksen2024open}, and ranking~\citep{robertson2009probabilistic}. Crawlers systematically discover and collect web content, which is then analyzed and organized into inverted indices for efficient retrieval. When users submit queries, search engines employ complex algorithms to assess document relevance and importance~\citep{fuhr1992probabilistic}. The PageRank algorithm revolutionized search by evaluating page authority based on the web's citation graph, determining a page's significance through its incoming links from other high-quality pages~\citep{page1999pagerank}. Modern search systems have further evolved to incorporate semantic understanding, user behavior data, and personalization to deliver increasingly relevant results~\citep{wang2024enhancing}. However, the contexts presented to the user may not always be relevant and accurate, due to the limited context for user-specific complex queries~\citep{leake2001towards} and the influence of advertisement bidding~\citep{linden2009pollution}.

\subsection{LLMs as Chatbots}

Leveraging recent advances in natural language processing (NLP) and hardware enhancements, large language models (LLMs) represent the latest evolution in information retrieval as user chatbots for tailored response generation. Models like ChatGPT~\cite{achiam2023gpt}, Claude~\cite{anthropic2023claude}, and LLaMa~\cite{touvron2023llama} transcend traditional static retrieval methods by engaging users through interactive dialogue with integrate and tailored solutions that not directly available online~\citep{thirunavukarasu2023large, kasneci2023chatgpt}.
% Unlike conventional search engines that require iterative user efforts in content browsing and query rewriting, LLM chatbots aggregate vast amounts of web-sourced knowledge within their parameters, effectively serving as compact representations of extensive online information~\citep{zhang2022opt, zeng2022glm}. 
Unlike conventional search engines that process each query independently—requiring iterative user efforts in providing contextual information and  content browsing, LLM chatbots maintain conversation history throughout interactions, Beyond that, they aggregate vast amounts of web-sourced knowledge within their parameters, effectively serving as compact representations of extensive online information~\citep{zhang2022opt, zeng2022glm}. 
Through supervised fine-tuning on conversation and instruction datasets~\citep{iyer2022opt}, coupled with reinforcement learning from human feedback (RLHF)~\citep{bai2022training}, these models optimize their responses for accuracy, relevance, and alignment with user preferences. Additionally, targeted prompt engineering~\citep{zhou2022large} and optimization techniques~\citep{li2021prefix}, alongside maintaining conversational context~\citep{callison2022dungeons}, further enhance the coherence and maturity of multi-turn interactions.

However, despite these advantages, relying exclusively on internal LLM knowledge presents notable challenges: (1) \textit{hallucinations}, where models generate plausible but inaccurate content~\citep{tam2023evaluating, yao2023llm, zhao2025dynamic}; (2) \textit{lack of awareness of recent events}, which compromises the timeliness of responses~\citep{chen2023can}; and (3) \textit{Limited context window}, hindering a comprehensive understanding of complex queries~\citep{wang2024news}. Therefore, integrating external information sources and employing advanced reasoning to verify retrieved data are crucial strategies for addressing these limitations, thus ensuring LLM chatbots deliver accurate, relevant, and up-to-date information~\citep{peng2023check}.

\subsection{LLMs with RAG} \label{sec: llm_rag}

To address the inherent limitations of LLMs mentioned above, particularly their static knowledge and tendency to hallucinations, Retrieval Augmented Generation (RAG) has emerged as a promising paradigm~\citep{prabhune2024deploying}. RAG integrates the generative capabilities of LLMs with retrieval systems to dynamically access relevant external information.
Early implementations of RAG primarily employed a straightforward "Retrieve-then-Read" workflow~\citep{ma2023query}, typically involving a single-step retrieval from a predefined local database or document collection. Although they improve upon purely parametric methods, such naive RAG systems can still struggle with inaccurate retrieval when faced with complex queries.

To solve this problem, multi-hop retrieval addresses the limitations of traditional single-hop retrieval by enabling iterative, sequential searches and reasoning steps across multiple data sources~\citep{jiang2023active}. Multi-hop retrieval incorporates iterative refinement, where intermediate retrieval outcomes guide subsequent queries, progressively building comprehensive context~\citep{zhang2025credible}. Although multi-hop retrieval has strong power generation capabilities, it also suffers from limitations due to the underlying techniques it employs. Early stage errors in reasoning paths can propagate throughout subsequent retrieval and reasoning steps, severely influencing the final output integrity~\citep{zhang2025credible}. Additionally, maintaining faithfulness to retrieved evidence poses ongoing difficulties, as language models frequently encounter conflicts between retrieved data and internal parametric knowledge~\citep{zheng2025multiple}.

\section{Towards Agentic Deep Research} \label{sec: research}

Many complex real-world problems, including open-domain question answering~\citep{yang2015wikiqa, chen2020open} and scientific discovery~\citep{lu2024ai, wang2023scientific, baek2024researchagent, schmidgall2025agent}, inherently require an iterative interplay between information retrieval and reasoning. A single search step often falls short of capturing comprehensive information, while isolated reasoning phases can fail to identify critical insights \citep{trivedi2023interleaving}.  By tightly integrating search and reasoning in a multi-step and interactive manner, these systems can progressively enhance the relevance and depth of retrieved knowledge and simultaneously refine the reasoning process underlying query interpretation, ultimately producing more accurate and contextually nuanced responses.
Here, reasoning actively influences search (e.g., refining search queries based on intermediate deductions), while retrieved information recursively refines reasoning in a dynamic feedback loop. Unlike the previous LLM with RAG framework in Section~\ref{sec: llm_rag}, where retrieval and reasoning occur in discrete and sequential stages, this approach treats them as interdependent, continuously co-evolving. 

This evolution in search methodologies gives rise to a transformative paradigm we define as Agentic Deep Research. In this paradigm, language models takes on the role of active information-seeking agents. Rather than a one-shot prompt + retrieve paradigm, an “agentic” LLM plans a series of steps: it can issue search queries, consult documents, browse on web, or even collaborate with other agents, all while refining its query understanding and response via iterative retrieval and reasoning. Inspired by the way human experts might research a question, we encapsulate this iterative synergy between \textit{reasoning} and \textit{search} in the term \textbf{Deep Research} highlighting its dynamic and interactive essence.
To substantiate our central position that LLM-driven Agentic Deep Research will inevitably become the predominant paradigm for future information-seeking---we ground our argument across three interlinked technical dimensions: reasoning capabilities as the foundation, principled approaches to incentivize search, and ecosystem-level momentum evidenced through benchmarks and implementations.

The evolution of reasoning capabilities in large language models represents a crucial stepping stone toward truly agentic systems, particularly in the context of deep research tasks. While Chain-of-Thought (CoT) prompting~\citep{wei2022chain} initially demonstrated the possibility of explicit reasoning processes, the real breakthrough lies in how reasoning mechanisms enable autonomous decision-making and strategic planning, essential for conducting deep research.
The transformation from simple CoT to more sophisticated reasoning frameworks marks a fundamental shift in how AI systems approach complex tasks. Rather than merely following predetermined patterns, modern reasoning frameworks enable systems to dynamically plan, execute, and adjust their approach based on intermediate outcomes. 
This capability is particularly evident in recent reinforcement learning-based optimization approaches~\citep{jaech2024openai,guo2025deepseek}, which have demonstrated unprecedented abilities in managing complex search tasks. These systems can autonomously determine when to initiate searches, formulate appropriate queries, and synthesize findings into coherent understanding, forming the cornerstone of agentic behavior.

The DeepSeek-R1~\citep{guo2025deepseek} represents a significant milestone in this evolution, demonstrating how reinforcement learning can optimize reasoning processes for complex mathematical tasks. By learning from experience and feedback, these systems develop sophisticated strategies for information gathering and synthesis, moving beyond simple pattern matching to true strategic planning. This advancement in reasoning capabilities provides the essential foundation for agentic deep research by enabling systems to autonomously evaluate information needs, strategically decompose complex queries, synthesize information across multiple sources while maintaining logical consistency, and adapt search strategies based on intermediate results and feedback \citep{jin2025search}.
These capabilities, rooted in advanced reasoning mechanisms, establish the preliminary foundation necessary for conducting deep research tasks that require strategic planning, iterative refinement, and complex decision-making. The integration of reinforcement learning with reasoning frameworks represents a crucial step toward truly agentic systems capable of conducting sophisticated research autonomously, marking a significant advancement from traditional search and retrieval paradigms \citep{chen2025learning}.

\subsection{Incentivizing Search with Reasoning Agents}

Within this paradigm, reasoning is not merely an auxiliary component applied post-retrieval; rather, it constitutes the core mechanism that determines when, what, and how to search \citep{wu2025agentic}. While prompting and supervised fine-tuning (SFT) serve as foundational techniques for instilling tool-use behaviors and basic query generation, they are inherently limited by their reliance on fixed instruction patterns and offline supervision \citep{wang2023towards, ghosh2024closer}. In contrast, reinforcement learning (RL) provides a principled framework for cultivating truly agentic behavior—enabling models to explore, self-correct, and adaptively optimize their retrieval strategies in open-ended, interactive environments~\citep{singh2025agentic, jin2025search, song2025r1}. This shift toward RL-incentivized search marks a critical step toward developing autonomous agents capable of reasoning-driven information acquisition.

\paragraph{Prompting and In-Context Learning: Bridging Search and Reasoning.}
Prompting methods have laid important groundwork for coupling reasoning with external information retrieval. ReAct~\citep{yao2023react} and its successors~\citep{li2025search, alzubi2025open} introduced paradigms where LLMs alternate between reasoning steps and tool use, guiding models to break problems down and issue relevant search queries mid-process. This enables iterative refinement of reasoning with retrieved evidence, improving factuality and coherence.
Extensions such as Search-o1~\citep{li2025search} and Open Deep Search (ODS)~\citep{alzubi2025open} prompt LLMs to actively consult web resources and integrate results into ongoing thought chains. In parallel, methods like Self-Ask~\citep{press2023measuring} and IRCoT~\citep{trivedi2023interleaving} embed search directly within step-by-step reasoning, generating sub-questions and retrieving partial answers in a recursive loop. 
These prompting approaches offer flexible templates to scaffold retrieval-enhanced reasoning. However, they rely on fixed prompting logic and do not provide incentives for exploring better search or reasoning paths, limiting their scalability for open-ended or high-stakes tasks.

\paragraph{Supervised Fine-Tuning: Hard-Coding Search Patterns.}
Supervised fine-tuning (SFT) takes a more structured approach by directly training LLMs on datasets that combine reasoning and retrieval. Toolformer~\citep{schick2023toolformer} and INTERS~\citep{zhu2024inters} illustrate how models can be trained to learn when and how to query external tools, assess retrieved information, and integrate it logically into final outputs.
SFT data typically comes from two sources: synthetic data generation (e.g., Toolformer~\citep{schick2023toolformer}, RAG-Studio~\citep{mao2024rag}) or instructional reformulation of existing datasets (e.g., INTERS~\citep{zhu2024inters}, InstructRetro~\citep{wang2024instructretro}). These enable LLMs to follow structured retrieval-reasoning sequences. However, such methods primarily encode static behaviors learned from data, not dynamic, adaptive behaviors optimized for diverse environments.
While prompting and SFT offer controlled environments for building retrieval-aware reasoning, they impose fixed search patterns and predefined goals. They do not equip agents with the ability to explore the open-ended, uncertain nature of real-world search tasks.

\paragraph{Reinforcement Learning: Optimizing Reasoning-Driven Search in the Wild.}
Reinforcement learning (RL) fundamentally changes the search paradigm by letting agents learn through trial and error in interactive environments. Instead of being told how to search, RL-trained agents are incentivized (through feedback or reward functions) to discover, refine, and adapt their reasoning and search strategies for specific goals.
Early systems like WebGPT~\citep{nakano2021webgpt} and RAG-RL~\citep{huang2025rag} demonstrated how reward signals (based on human feedback or factual correctness) can guide multi-step retrieval policies that improve response accuracy and trustworthiness. More modular designs like M-RAG~\citep{wang2024m} separate reasoning and retrieval into specialized agents, each trained to collaborate via shared RL objectives.
Recent RL-based systems such as Search-R1~\citep{jin2025search}, R1-Searcher~\citep{song2025r1}, DeepResearcher~\citep{zheng2025deepresearcher}, ZeroSearch~\citep{sun2025zerosearch}, and WebAgent-R1~\citep{wei2025webagent} operate in various search environments from static local corpora and open search APIs to real-world web interfaces. These agents learn to decompose complex tasks, plan query sequences, verify evidence, and adjust their strategies based on environment feedback. Such behaviors are difficult to teach through SFT or prompts alone.
Importantly, ReSearch~\citep{chen2025learning} and ReARTeR~\citep{sun2025rearter} go a step further by optimizing not just factual correctness, but also alignment with transparent, interpretable reasoning. ReARTeR introduces a dual-model approach that incentivizes both outcome quality and step-wise explainability, offering a more human-aligned path to trustworthy automation.

\subsection{Benchmarks and Open-Source Implementations:} 

\begin{figure*}[!th]
  \centering
    \includegraphics[width=0.85\textwidth]{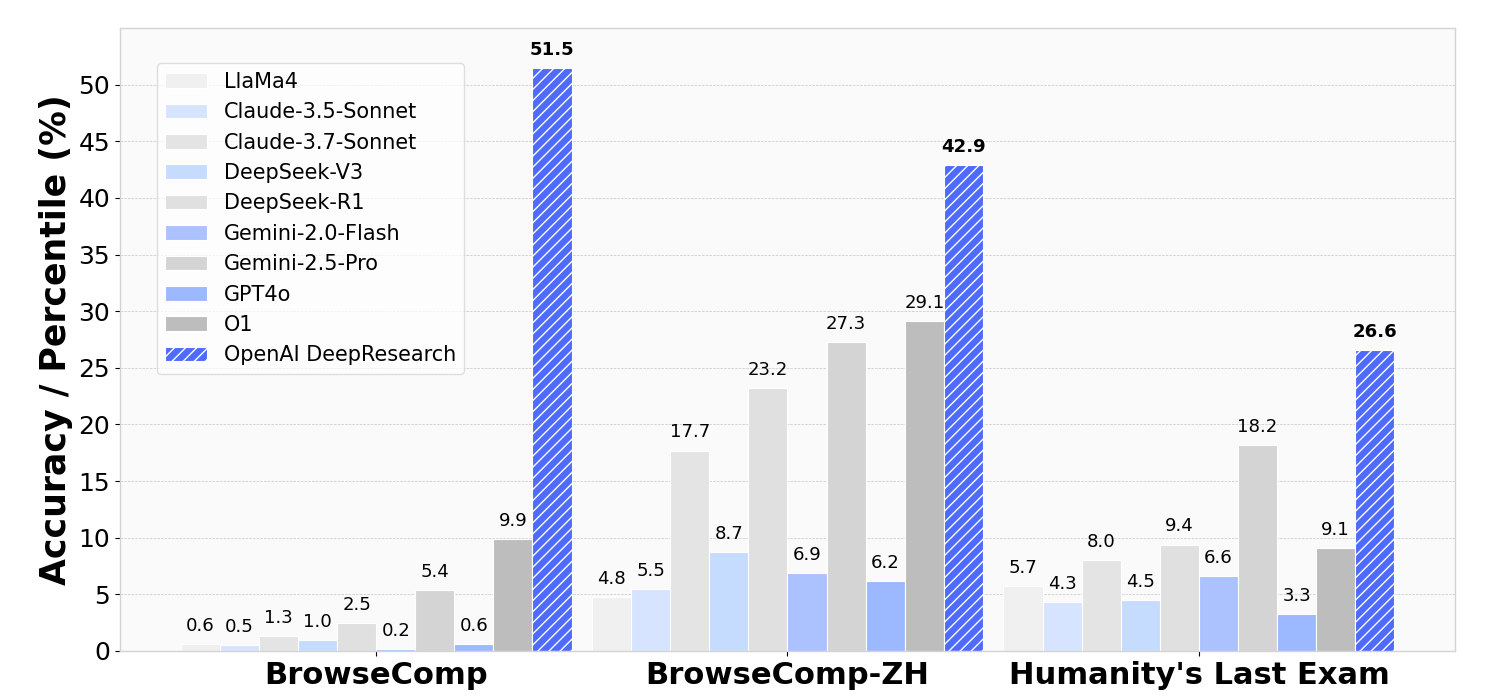}
  \caption{Benchmarks of 5 standard LLMs, 4 reasoning LLMs, and 1 agentic deep research model (OpenAI Deep Research) on BrowseComp, BrowseComp-ZH, and Humanity's Last Exam.}
  \label{fig: Bench}
\end{figure*}

\paragraph{Deep Research Benchmarks}
To rigorously compare the capabilities of standard LLMs, reasoning LLMs, and agentic deep research models in realistic and high-stakes scenarios, we adopt and evaluate three representative benchmarks, including BrowseComp~\citep{wei2025browsecomp}, BrowseComp-ZH~\citep{zhou2025browsecomp}, and Humanity’s Last Exam (HLE)~\citep{phan2025humanity}, each targeting distinct dimensions of agentic deep research. BrowseComp assesses an agent’s ability to conduct multi-step, open-ended web searches to retrieve non-trivial information, while BrowseComp-ZH extends this challenge to the Chinese web, introducing additional linguistic complexity. In contrast, HLE focuses on presenting expert-level questions across diverse academic domains that cannot be solved through naive retrieval alone. These tasks require agents to synthesize evidence from obscure or fragmented sources (e.g., identifying policy changes from regional Chinese documents or resolving historical ambiguities) or to reason through abstract academic problems, where more details can be found in Appendix~\ref{app: bench}. As shown in Figure 2, standard LLMs perform poorly across these benchmarks—typically below 10\% on BrowseComp datasets and under 20\% on HLE. In comparison, the OpenAI Deep Research agent achieves significantly higher scores—51.5\% on BrowseComp, 42.9\% on BrowseComp-ZH, and 26.6\% on HLE—demonstrating the effectiveness of reasoning-integrated search in advancing the frontier of intelligent information-seeking systems.

\paragraph{Open-Source Implementations}

To empirically ground the rising momentum behind Agentic Deep Research, we examine GitHub star trajectories for recent open-source implementations within this paradigm and we provide detailed information in Appendix~\ref{app: github}. After excluding the two most-starred repositories (to mitigate skew from viral or legacy projects) and the two least-starred (to reduce statistical noise), we observe a clear upward trajectory across nearly all remaining projects since early 2025. Notably, \textit{deep-searcher} and \textit{deer-flow} experienced rapid surges, reaching thousands of stars within weeks. Even smaller-scale efforts, including \textit{DeepResearcher} and \textit{R1-Searcher}, display a consistent upward trend, highlighting the breadth of innovation within the agentic search space. These patterns, along with the average Github star trends, indicate not only a technical transition but also a broader cultural and developmental shift: the open-source community is increasingly converging around reasoning-driven, agentic deep research as a leading framework for information seekin. This empirical momentum reinforces our position statement—that Agentic Deep Research LLM-cored Agentic Deep Research framework will inevitable become the dominant paradigm for future information search.

\begin{figure}[htbp]
  \centering
  \begin{subfigure}[t]{0.5\textwidth}
    \centering
    \includegraphics[width=0.85\textwidth]{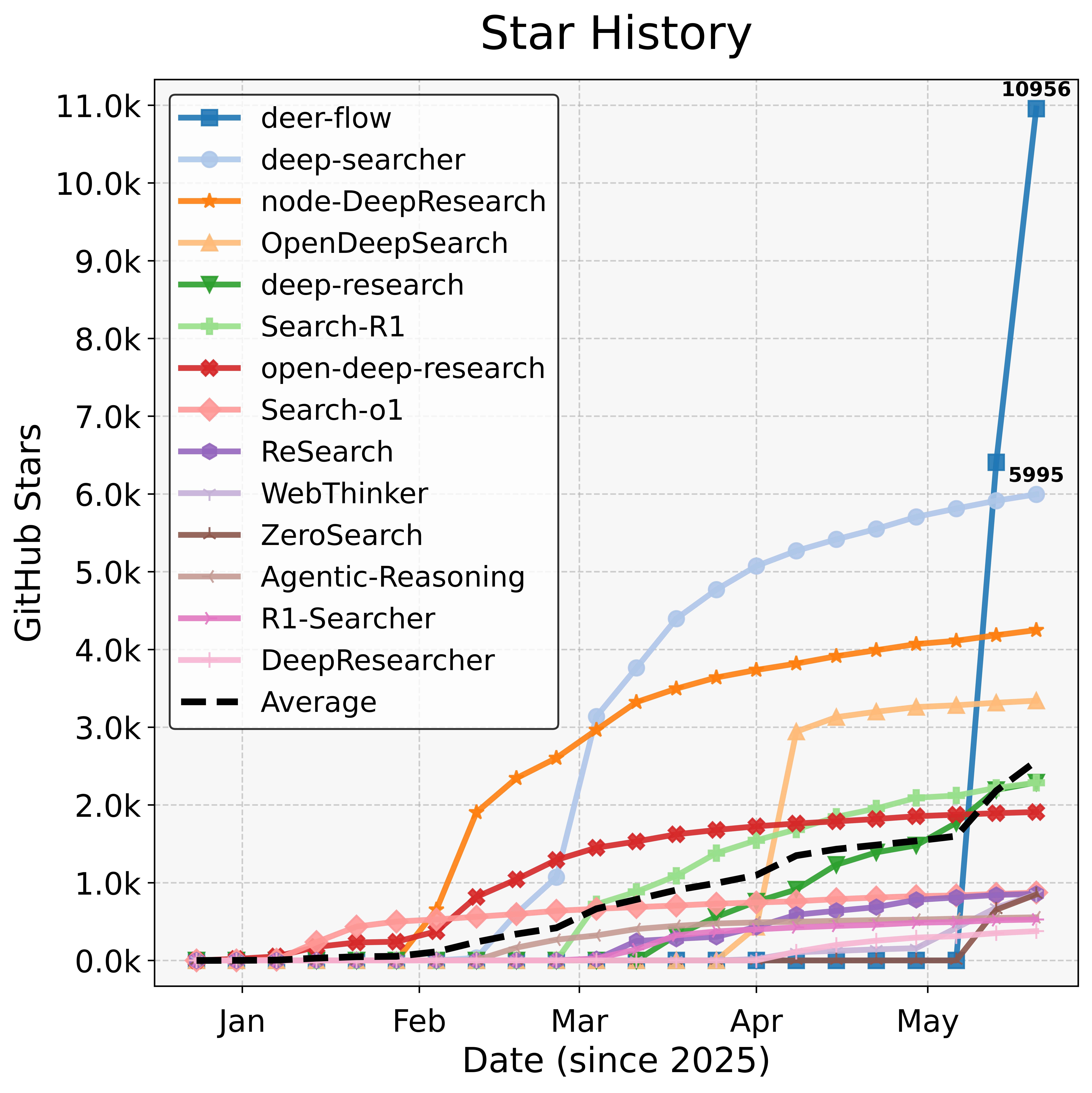}
    \caption{GitHub star trend for open-source repositories.}
    \label{fig:github}
  \end{subfigure}
  \hfill
  \begin{subfigure}[t]{0.49\textwidth}
    \includegraphics[width=\textwidth]{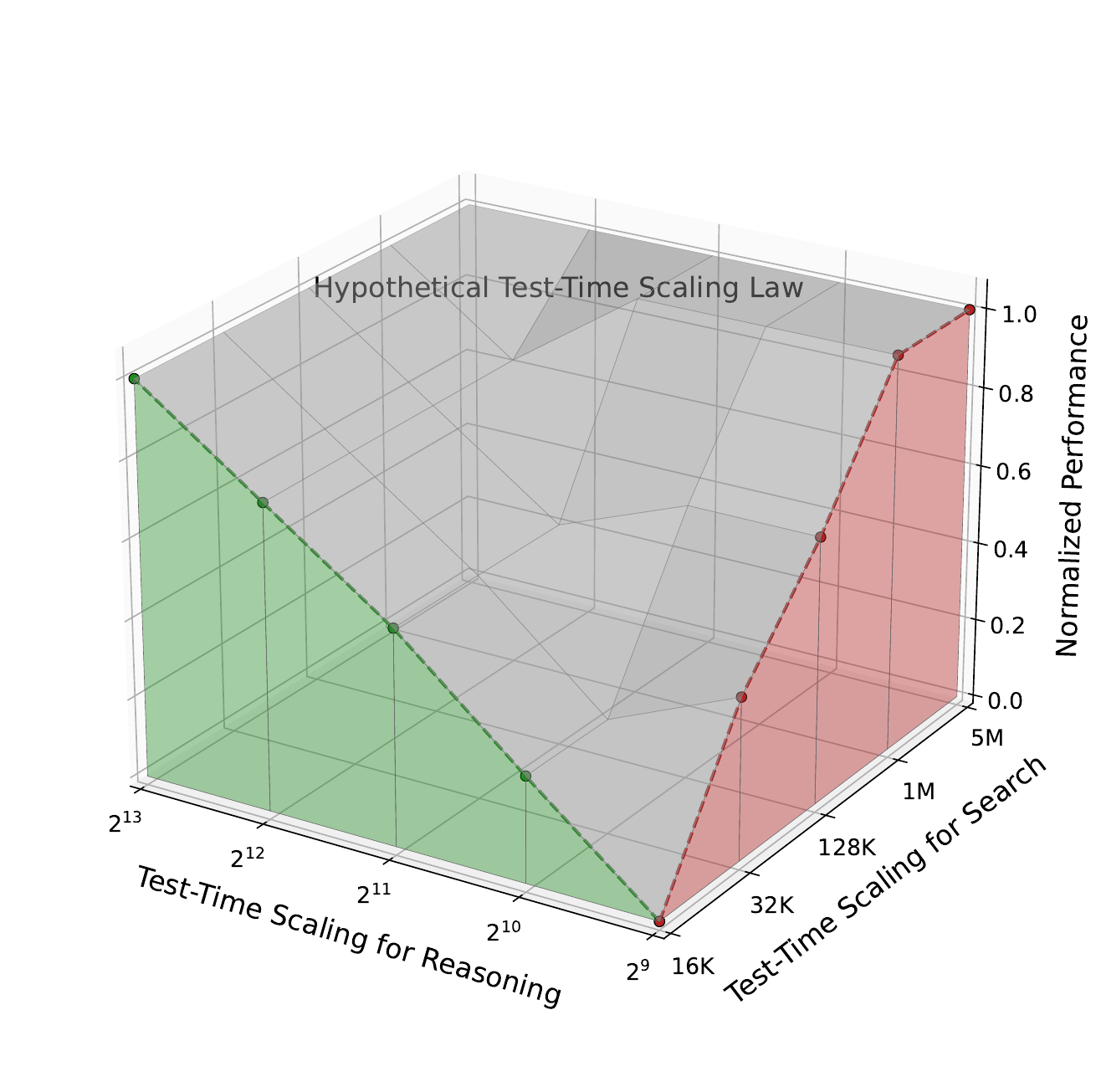}
  \caption{{Test-time scaling law for agentic deep research.}}
  \label{fig:tts}
  \end{subfigure}
  \caption{(a) Open-source star trend for agentic deep research. {(star counts recorded up to 22 May 2025)}. (b) Test-time scaling (TTS) law for agentic deep research, where data before performance normalization for reasoning TTS is from S1 on AIME24~\citep{muennighoff2025s1} and for search TTS is from IterDRAG on MuSiQue~\citep{yueinference}. The TTS is measured in inference tokens.}
  \label{fig:parallel}
\end{figure}

\section{Test-Time Scaling Law for Deep Research}
Building upon the stated position, we introduce the Test-Time Scaling (TTS) law for Agentic Deep Research, an hypothesis predicting the performance improvements achievable through extended computational resources during inference. Figure~\ref{fig:tts} illustrates this hypothesis, normalized performance improve linearly through scaling of internal reasoning depth and external knowledge exploration. Data supporting this observation comes from two representative evaluations: S1 on the AIME24 dataset \citep{muennighoff2025s1}, which tests reasoning-based scaling on advanced multi-step mathematical reasoning problems, and IterDRAG on MuSiQue \citep{yueinference}, which focuses on search-based scaling via multi-hop retrieval tasks. As in Figure~\ref{fig:tts}, the diagonal plane connecting empirical data points interpolated in the three-dimensional plot represents our hypothetical TTS for Agentic Deep Research. Tasks requiring deeper internal knowledge utilization (reasoning), such as solving complex math problems and logic puzzles, are better aligned with the green plane, whereas tasks demanding extensive external knowledge exploration (search), like medical QA, will align more closely with the red plane. Here, we illustrate the TTS law and trade-offs along the search and reasoning axes.

\subsection{Reasoning --- TTS of Internal Knowledge Utilization}
Recent studies have shown that LLMs do more than recall memorized facts or patterns. When given extra computational resource to ``think'', they can also perform deeper reasoning steps. This pattern, known as the test-time scaling law for reasoning, indicates that as a model takes more inference steps, including building longer chains of thought \citep{chen2025towards}, iterative self-refinement \citep{madaan2023self}, or self-consistency decoding \citep{hao2023reasoning}, its accuracy in complex tasks steadily improves.
This phenomenon also suggests that the internal knowledge embedded in LLMs is not fully exposed in a single forward pass. Instead, deeper understanding can be progressively uncovered through extended inference. 
% In tasks like multi-hop QA or mathematical problem solving, performance has been observed to scale with reasoning depth, rather than model size alone.
In the context of Agentic Deep Research, this highlights a key shift from producing one-shot answers to engaging in active, multi-step reasoning. Instead of relying solely on external retrieval or prompting tricks, the model's internal knowledge becomes a reusable and expandable resource. This inference-time flexibility is crucial for handling complex, open-ended queries, and positions reasoning as a core axis of scalable capability in LLM-based research agents~\citep{huang2024understanding}.

\subsection{Search --- TTS of External Knowledge Exploration}
In addition to making full use of the internal knowledge of LLMs, exploring abundant external knowledge effectively is another key to achieving TTS in Agentic Deep Research systems. When performing external knowledge search, it is often difficult to obtain all the important information in a single-step retrieval~\citep{shao2023enhancing,jiang2024retrieve}. To overcome this bottleneck, consistent with the agentic RAG insights, existing work mainly explores the scaling potential of the search phase through \textit{iterative} search and \textit{long-context} RAG. Representatively, many works have explored improving the test-time retrieval performance through iterative RAG/search, which introduces a dynamic and multi-step retrieval to knowledge search via task decomposition~\citep{trivedi2023interleaving,asai2023self,xiong2024improving}. They generally show that iterative multi-step retrieval under \textit{a proper number of iterations} can also enhance RAG's performance. \citet{yueinference} recently further observed an important experimental phenomenon that under optimal inference parameters, the performance improves nearly linearly with increasing test-time computation. 
The gradual expansion of knowledge retrieval {from split local text chunks towards an increasingly precise retrieval executed within nearly global external knowledge bases} would become a more powerful development trend in the test-time search process.

\subsection{TTS Trade-offs for Search and Reasoning}
 Agentic Deep Research systems integrate both search and reasoning, each consuming part of a limited token budget. Under such constraints, a natural trade-off emerges: allocating more tokens to search (e.g., issuing broader or more detailed queries) reduces the capacity available for reasoning (e.g., multi-hop inference or synthesis), and vice versa. This balance is task-dependent. Search-heavy tasks such as multi-hop RAG~\citep{xiong2024improving,shao2023enhancing} or literature surveys prioritize broader content access, while reasoning-heavy tasks like causal analysis or math verification~\citep{snell2024scaling} require deeper internal processing. Building systems that adaptively allocate token budgets between search and reasoning based on task characteristics is critical for maximizing effectiveness and efficiency in Agentic Deep Research. We anticipate the emergence of a test-time scaling law that governs the optimal balance between search and reasoning under different task conditions. Furthermore, training models that can dynamically allocate and manage budget across these two components is a defining capability of next-generation deep research systems.

\section{Alternative View and Discussions}

\paragraph{Retaining Human Primacy in Search}
In contrast to the position advanced in this paper that the future of information seeking paradigm is LLM-driven Agentic Deep Research. An alternative viewpoint contends that search should remain fundamentally a human-led activity, with artificial intelligence systems serving primarily as assistive, not autonomous, tools. This counter-position emphasizes that human with primary involvement is indispensable for ensuring trust, interpretability, and epistemic responsibility in open-ended inquiry tasks \citep{mehrotra2024systematic}.
First, despite the recent progress in agentic reasoning capabilities, autonomous systems still lack robust models of user intent, contextual nuance, and domain-specific ethics. These elements are often essential in complex or high-stakes search scenarios such as scientific research, legal interpretation, or public policy analysis. Advocates of human-led search argue that delegating the full pipeline (from information retrieval to reasoning and synthesis) to LLM agents risks introducing misaligned conclusions, opaque decision paths, and reduced user oversight.
In addition, from a trust and accountability perspective, human-directed systems afford greater transparency and traceability. While autonomous agents can produce fluent and plausible outputs, they also introduce increased risk of hallucinations, spurious correlations, or unjustified reasoning steps. Maintaining human control over search allows users to apply critical judgment, verify information provenance, and assume responsibility for downstream decisions, a particularly salient concern in regulated or high-risk domains ~\citep{zou2025survey, zou2025call}.

\section{Open Problems and Future Opportunities}

\paragraph{Human-in-Loop and Trustworthy}
This suggests that progress in AI search systems should prioritize augmenting human capabilities rather than replacing them. This includes building interfaces that support iterative refinement, expose intermediate reasoning steps, and enable meaningful user feedback. In this framing, AI functions less as an autonomous researcher and more as a powerful assistant embedded within a human-centered workflow~\citep{shneiderman2022human}. 
To build trustworthy Deep Research systems, human interactions play an important role. Several important research questions need to be solved.
\textit{(1) Search Content Access Control.} Implementing fine-grained access control mechanisms ensures that users can only access information appropriate to their roles and permissions. This is particularly important in domains where role-based access to information is essential for privacy, regulatory compliance, or competitive confidentiality.
\textit{(2) Human Verification and Feedback Mechanisms.} Incorporating human oversight at critical stages of the AI search process can significantly improve the accuracy and reliability of the system~\citep{zou2025survey, zou2025call}. Designing systems that facilitate user feedback enables continuous learning and adaptation, aligning AI outputs more closely with user expectations~\citep{zhong2024memorybank, zou2025call} and user-specific requirements~\citep{zhang2025personaagent}.

\paragraph{In-Domain Expert-Level Deep Research}
Deep Research is emerging as the new internet traffic entrance and current Deep Research systems focus on general-purpose domains. However, in highly specialized fields such as medicine, law, and biology, in-domain expert-level deep research is essential to ensure accuracy, relevance, and usability. Domain-specific research poses unique challenges that general-purpose agents cannot meet. We highlight three key research directions:
 \textit{(1) Domain-Specific Database Construction.}  
In many fields, data is scattered across fragmented databases with inconsistent coverage and interfaces—e.g., in bioinformatics~\citep{benson2012genbank,uniprot2018uniprot,dyer2025ensembl} or medicine~\citep{knox2024drugbank,landrum2018clinvar}. This fragmentation hampers retrieval and reasoning. Future systems must build unified, structured, and queryable domain-specific databases to enable effective deep research.
 \textit{(2) Domain-Grounded Reasoning.}  
Reasoning paradigms vary by field: legal reasoning relies on precedent, medical reasoning on diagnostic codes, scientific reasoning on hypothesis testing. Agentic systems must align with these paradigms, adapting their planning and inference mechanisms to domain-specific logic and workflows.
%  {(3) Domain-Aware Prompt Optimization.}  
% Users often lack the technical vocabulary to formulate effective queries in specialized fields~\citep{wang2022automated}. For example, meaningful prompts in healthcare may require ICD-10 codes to specify diseases. Future systems should include prompt optimizers that translate lay queries into structured, domain-relevant inputs, bridging the gap between user intent and expert knowledge formats.

\paragraph{Structure-Organized Deep Research Systems} Structured (graph) data, composed of nodes (representing entities) and edges (representing relationships), offers an intuitive and systematic way to represent complex relationships and knowledge associations~\citep{hamilton2017inductive,veličković2018graph}. This structured organization can significantly enhance the deep research process in the following ways: \textit{(1) Iterative Data Organization}: There are many intermediate searching and reasoning contents during agentic retrieval and reasoning. Structuring these iterative contents helps agents maintain coherence and relevance in the long context~\citep{li2024graphreader}. As structured data clarifies knowledge relationships, agents can follow these links effectively when generating answers. This allows them to find relevant content more clearly and avoid contradictions or illogical situations. \textit{(2) Multi-Agent Deep Research}: Effective task allocation and coordination are essential in multi-agent collaboration~\citep{luo2025large}. Graph structure helps agents understand task requirements based on their roles and relationships~\citep{zhuge2024gptswarm,zhang2024g}. 
Further, in multi-agent collaboration, timely and accurate information exchange is vital. Graph data serves as an effective information carrier, allowing agents to structure complex information.
Learning (message passing) from this structured data enhances efficient information sharing among agents based on their relationships.

\paragraph{From Textual Space to Multi-Modality} For Agentic Deep Research systems to truly emulate human research capabilities, they must transcend textual limitations and integrate diverse information modalities including images, audio-visual content, and structured data. Different modalities inherently encode distinct knowledge types: text conveys abstract concepts and logical relationships, images provide visual instances and spatial information, while videos capture temporal dynamics and sequential processes. This evolution requires a qualitative cognitive leap in knowledge integration rather than merely expanding input channels. Critical research directions include:  \textit{(1) Cross-modal Semantic Alignment:} deep cross-modal semantic alignment within unified representational spaces that support context-aware reasoning across modalities~\citep{li2025vista};  \textit{(2) Information Fusion and Conflict Resolution:} robust information fusion with sophisticated conflict resolution mechanisms when modalities present inconsistent information~\citep{bi2025reasoning}; and  \textit{(3) Multi-modal Knowledge Acquisition:} efficient multi-modal knowledge retrieval and comprehension from heterogeneous, large-scale data repositories~\citep{abootorabi2025ask}.

\paragraph{Efficient Test-Time Scaling} As Agentic Deep Research systems advance, operational efficiency and cost management become critical scaling determinants. Optimization must target both reasoning and search:  \textit{(1) Reasoning Efficiency}, balancing computational depth against resource utilization through techniques like capability transfer to smaller models or latent reasoning approaches~\citep{sui2025stop}. Such efficiency improvements potentially extend beyond mere compression of reasoning chains to manifest as higher-order intelligence through \textit{resource self-management}~\citep{li2025selfbudgeter}. 
\textit{(2) Search Efficiency}: Optimizing search beyond naive retrieval requires addressing scalable retrieval from massive heterogeneous sources, efficient long-context processing~\citep{luo2025does}, and iterative search with intelligent query refinement and adaptive stopping criteria~\citep{singh2025agentic}. Future systems must implement budget-aware strategies that dynamically adjust search workflows.

\section{Conclusion}

This paper has articulated a compelling trajectory from traditional web search paradigms towards the inevitable ascendancy of Agentic Deep Research. By systematically addressing the limitations of prior search engine systems and emphasizing the transformative potential of iterative reasoning and search enabled through advanced reinforcement learning frameworks, we demonstrate that agentic systems significantly outperform traditional models across complex benchmarks. The empirical trends observed in both academic evaluations and open-source implementations reinforce this shift, indicating broad recognition and adoption within different communities. Nevertheless, recognizing legitimate concerns regarding human oversight and transparency, future developments must incorporate hybrid frameworks that optimize both autonomous agentic capabilities and human-in-the-loop interactions. In proposing multiple open challenges and opportunities, we foresee Agentic Deep Research as not only the dominant paradigm but also a profoundly human-centered LLM-driven advancement in knowledge acquisition and synthesis.

% \begin{ack}
% Use unnumbered first level headings for the acknowledgments. All acknowledgments

% Do {\bf not} include this section in the anonymized submission, only in the final paper. You can use the \texttt{ack} environment provided in the style file to automatically hide this section in the anonymized submission.
% \end{ack}

\bibliographystyle{apalike}
\bibliography{main.bib}

%%%%%%%%%%%%%%%%%%%%%%%%%%%%%%%%%%%%%%%%%%%%%%%%%%%%%%%%%%%%

\appendix

\section{Deep ReSearch Benchmarks}\label{app: bench}

\paragraph{BrowseComp}~\citep{wei2025browsecomp}
is a benchmark introduced by OpenAI to evaluate the capabilities of AI agents in conducting complex web searches. Comprising 1,266 challenging questions, it assesses an agent's ability to persistently navigate the internet to locate hard-to-find, entangled information. Unlike traditional benchmarks that focus on retrieving easily accessible facts, BrowseComp emphasizes tasks where answers are deliberately obscured, requiring agents to demonstrate advanced reasoning, strategic search planning, and adaptability. The benchmark's design ensures that while answers are difficult to discover, they are straightforward to verify, facilitating reliable evaluation of agent performance. BrowseComp serves as a critical tool for advancing research in developing AI systems capable of sophisticated information retrieval and reasoning across the web.

\paragraph{BrowseComp-ZH}~\citep{zhou2025browsecomp}
is a high-difficulty benchmark developed to evaluate the web browsing and reasoning capabilities of large language models (LLMs) within the Chinese information ecosystem. Recognizing that existing benchmarks like BrowseComp focus primarily on English-language contexts, BrowseComp-ZH addresses the unique challenges posed by the Chinese web, including linguistic complexity, fragmented information across diverse platforms, and varying search engine infrastructures.
The benchmark comprises 289 multi-hop questions spanning 11 diverse domains such as film, art, medicine, geography, history, and technology. Each question is meticulously reverse-engineered from a concise, objective, and easily verifiable answer (e.g., a date, number, or proper noun). A two-stage quality control protocol ensures high question difficulty and answer uniqueness. Notably, the questions are designed so that answers are not readily retrievable via standard search engines, requiring models to engage in complex reasoning and information synthesis.

\paragraph{Humanity's Last Exam (HLE)}~\citep{phan2025humanity} is a multi-modal benchmark designed to evaluate the reasoning and problem-solving capabilities of large language models (LLMs) across a broad spectrum of academic disciplines. Developed collaboratively by the Center for AI Safety and Scale AI, HLE comprises 3,000 expert-crafted questions spanning mathematics, humanities, natural sciences, and more.
Unlike BrowseComp and BrowseComp-ZH, where agents can locate and answer questions by analyzing information retrieved from the web, HLE presents 'closed‑book' academic challenges (the answers aren’t directly available online) that demand deep reasoning and specialized domain expertise, going well beyond what surface‑level online searches can support. Each question in HLE is designed to be unambiguous and verifiable, yet not readily answerable through internet search, thereby testing the intrinsic reasoning abilities of LLMs.

\section{Open-Source Deep Research Implementations} \label{app: github}

\renewcommand{\arraystretch}{1.4}

% === Color definitions ===
\definecolor{header}{RGB}{217,225,242}

% === Column types & spacing ===
\newcolumntype{T}[1]{>{\bfseries\raggedright\arraybackslash}p{#1}} % Bold raggedright
\newcolumntype{X}{>{\raggedright\arraybackslash}X}             % Flexible width
\newcolumntype{C}[1]{>{\centering\arraybackslash}p{#1}}       % Centered
\setlength{\tabcolsep}{0.1pt}

\begin{table*}[ht]
  \centering
  \fontsize{6pt}{6pt}\selectfont
  \caption{Overview of open-source deep research implementations.}
  \begin{tabularx}{\textwidth}{@{}%
    T{3.5cm}  % Name
    C{2.0cm}  % Base Model
    C{1.5cm}    % Optimization
    % C{2cm}    % Agent Architecture
    C{2.5cm}  % Train Data
    C{3.5cm}  % Evaluation Data
    C{0.8cm}  % Link
  @{} }
    \toprule
    \rowcolor{header}
    \textbf{Name} & \textbf{Base Model} & \textbf{Optimization} & \textbf{Training Data} & \textbf{Evaluation Data} & \textbf{Link} \\
    \midrule
    %—— Centralized ——
    Agentic Reasoning \citep{wu2025agentic}
      & N/A
      & Prompt
      & N/A
      & GPQA
      & \href{https://github.com/theworldofagents/Agentic-Reasoning}{Link} \\

    Search-o1 \citep{li2025search}
      & Qwen
      & Prompt
      & N/A
      & GPQA, MATH500, AMC2023, AIME2024, LiveCodeBench, Natural Questions, TriviaQA, HotpotQA, 2Wiki, MuSiQue, Bamboogle
      & \href{https://github.com/sunnynexus/Search-o1}{Link} \\

    Open Deep Search \citep{alzubi2025open}
      & DeepSeek, Llama
      & Prompt
      & N/A
      & SimpleQA, FRAME
      & \href{https://github.com/sentient-agi/OpenDeepSearch}{Link} \\
      
    Search-R1 \citep{jin2025search}
      & Llama, Qwen
      & RL
      & NQ, HotpotQA
      & NQ, TriviaQA, PopQA, HotpotQA, 2WikiMultiHopQA, MuSiQue, Bamboogle
      & \href{https://github.com/PeterGriffinJin/Search-R1}{Link} \\

    DeepResearcher \citep{zheng2025deepresearcher}
      & Qwen
      & RL
      & NQ, TQ, HotpotQA, 2WikiMultiHopQA
      & MuSiQue, Bamboogle, PopQA, NQ, TQ, HotpotQA, 2WikiMultiHopQA
      & \href{https://github.com/GAIR-NLP/DeepResearcher}{Link} \\

    R1-Searcher \citep{song2025r1}
      & Llama, Qwen
      & RL
      & HotpotQA, 2WikiMultiHopQA
      & HotpotQA, 2WikiMultiHopQA, MuSiQue, Bamboogle
      & \href{https://github.com/RUCAIBox/R1-Searcher}{Link} \\

    ReSearch \citep{chen2025research}
      & Qwen
      & RL
      & MuSiQue
      & HotpotQA, 2WikiMultiHopQA, MuSiQue, Bamboogle
      & \href{https://github.com/Agent-RL/ReSearch}{Link} \\

    ZeroSearch \citep{sun2025zerosearch}
      & Llama, Qwen
      & RL
      & NQ, HotpotQA
      & NQ, TriviaQA, PopQA, HotpotQA, 2WikiMultiHopQA, MuSiQue, Bamboogle
      & \href{https://github.com/Alibaba-NLP/ZeroSearch}{Link} \\

    IKEA \citep{huang2025reinforced}
      & Qwen
      & RL
      & NQ, HotpotQA
      & NQ, HotpotQA, PopQA, 2Wikimultihopqa
      & \href{https://github.com/hzy312/knowledge-r1}{Link} \\
      
    Webthinker \citep{li2025webthinker}
      & Qwen
      & RL
      & SuperGPQA, WebWalkerQA, OpenThoughts, NaturalReasoning, NuminaMath
      & GPQA, GAIA, WebWalkerQA, Humanity’s Last Exam
      & \href{https://github.com/RUC-NLPIR/WebThinker}{Link} \\
      
    gpt-researcher
      & OpenAI Series
      & Prompt
      & N/A
      & N/A
      & \href{https://github.com/assafelovic/gpt-researcher}{Link} \\

    deep-searcher
      & DeepSeek, OpenAI Series, Claude, Gemini, Grok, Qwen, Llama, GLM
      & Prompt
      & N/A
      & N/A
      & \href{https://github.com/zilliztech/deep-searcher}{Link} \\

    nanoDeepResearch
      & OpenAI Series, Claude
      & Prompt
      & N/A
      & N/A
      & \href{https://github.com/liyuan24/nanoDeepResearch}{Link} \\

    \addlinespace
    %—— Decentralized ——
    DeerFlow
      & OpenAI Series, Qwen
      & Prompt
      & N/A
      & N/A
      & \href{https://github.com/bytedance/deer-flow}{Link} \\

    \addlinespace
    %—— Single ——
    deep-research
      & DeepSeek, OpenAI Series
      & Prompt
      & N/A
      & N/A
      & \href{https://github.com/dzhng/deep-research}{Link} \\

    open-deep-research
      & OpenAI Series, DeepSeek, Claude, Gemini
      & Prompt
      & N/A
      & N/A
      & \href{https://github.com/btahir/open-deep-research}{Link} \\

    r1-reasoning-rag
      & DeepSeek
      & Prompt
      & N/A
      & N/A
      & \href{https://github.com/deansaco/r1-reasoning-rag}{Link} \\

    node-DeepResearch
      & Gemini, OpenAI Series
      & Prompt
      & N/A
      & N/A
      & \href{https://github.com/jina-ai/node-DeepResearch}{Link} \\

    deep-research
      & Gemini, OpenAI Series, DeepSeek, Claude, Grok
      & Prompt
      & N/A
      & N/A
      & \href{https://github.com/u14app/deep-research}{Link} \\
    \bottomrule
  \end{tabularx}
  \label{tab:agentic-frameworks}
\end{table*}
% \section{Technical Appendices and Supplementary Material}
% Technical appendices with additional results, figures, graphs and proofs may be submitted with the paper submission before the full submission deadline (see above), or as a separate PDF in the ZIP file below before the supplementary material deadline. There is no page limit for the technical appendices.

%%%%%%%%%%%%%%%%%%%%%%%%%%%%%%%%%%%%%%%%%%%%%%%%%%%%%%%%%%%%

\end{document}